\pdfoutput=1
\documentclass{aa}
\usepackage[varg]{txfonts}
\usepackage{natbib, amsmath, amssymb, amsfonts, graphicx}
\usepackage{mathtools}
\usepackage{caption}
\usepackage{subcaption}
\usepackage{comment}
\usepackage{xcolor}
\usepackage{diagbox} 
\usepackage[squaren, Gray, cdot]{SIunits}
\usepackage{hyperref}
\hypersetup{colorlinks=true,allcolors=black,citecolor=blue, linkcolor=blue}
\usepackage{bigints}
\usepackage{lscape}
\usepackage{fancyvrb}

\RecustomVerbatimCommand{\VerbatimInput}{VerbatimInput}%
{fontsize=\footnotesize,
 frame=lines,  
 framesep=2em, 
 rulecolor=\color{gray},
 commandchars=\|\(\), 
 commentchar=*        
}

\usepackage{breqn} 
\everymath{\displaystyle}

\def \B{{\boldsymbol{B}}}

\def\et{{\boldsymbol{e_{\theta}}}}
\def \er{{\boldsymbol{e_{r}}}}
\def\ep{{\boldsymbol{e_{\varphi}}}}

\def \di{\boldsymbol{\nabla} \cdot}
\def \nab{\boldsymbol{\nabla}}
\def \rot{\boldsymbol{\nabla} \times}
\newcommand{\adv}[2]{\ensuremath \left(\boldsymbol{#1}\cdot \boldsymbol{\nabla}\right) #2}

\newcommand{\Dt}[1]{\ensuremath\frac{\partial #1}{\partial t}}

\title{Stochastic excitation of waves in magnetic stars}
\subtitle{I. Scaling laws for the modes amplitudes}
\author{L. Bessila \inst{1}
\and S. Mathis \inst{1}
}
\date{Received XX; accepted YY}

\institute{Université Paris-Saclay, Université de Paris, CEA, CNRS, Astrophysique, Instrumentation et Modélisation Paris-Saclay,  F-91191, Gif-sur-Yvette, France
\\\email{leila.bessila@cea.fr}}

\titlerunning{Stochastic excitation of waves in magnetic stars}
\authorrunning{Bessila, L. \& Mathis, S.}

\abstract{Stellar oscillations are key to unravelling stars' properties, such as their mass, radius and age. In turn, it enables to date and characterize their exoplanetary systems. Amplitudes of acoustic (p-) modes in solar-like stars are intrinsically linked to their convective turbulent excitation source, which in turn is influenced by magnetism. 
In the observations of the Sun and stars, the amplitudes of the modes are modulated following their magnetic activity cycles: the higher the magnetic field, the lower the modes' amplitudes. When the magnetic field is strong, it can even inhibit acoustic modes, which are not detected in a majority of solar-like stars presenting a strong magnetic activity. Magnetic fields are known to freeze convection when stronger than a critical value: an "on-off" approach is used in the literature.}{In this work, we investigate the impact of magnetic fields on the stochastic excitation of acoustic modes.}
{First, we generalise the forced wave equation formalism, including the effects of magnetic fields. Second, we assess how convection is affected by magnetic fields using results from Magnetic Mixing-Length Theory.}
{We provide the source terms of the stochastic excitation, including a new magnetic source term and the Reynolds stresses. We scaling laws for the amplitudes of the modes, taking into account both the driving and the damping. Those scalings are based on the inverse Alfvén dimensionless parameter: the damping increases with the magnetic field, and reaches a saturation threshold when the magnetic field is strong. The driving of the modes diminishes when the magnetic field becomes stronger, the turbulent convection being weaker.}
{As expected from the observations, we find that a higher magnetic field diminishes the resulting modes amplitudes. Evaluating the inverse Alfvén number in stellar models provides a means to estimate the expected amplitudes of acoustic modes in magnetic active solar-type stars.}


\begin{document}
\maketitle

\section{Introduction}
\par Over the past decades, studying the oscillation  modes of the Sun and stars allowed us to make transformative progresses in understanding stellar physics. \cite{ulrich_five-minute_1970} and \cite{leibacher_new_1971} were the first to identify the solar five-minute oscillations as acoustic p modes. Knowing the acoustic oscillation frequencies, space-based helio, and asteroseismology became a powerful method to constrain the global parameters of stars, such as their mass, radius, and age \citep{christensen-dalsgaard_bright_2015, garcia_asteroseismology_2019, aerts_probing_2021}. Combined with the transit and the radial velocity methods this allows to characterise exoplanets and the dynamics of their host star \citep{huber_planetary_2009, gizon_seismic_2013}.
Moreover, stellar seismology has enabled us to probe the structure and the internal rotation of stars \citep[e.g.][]{thompson_internal_2003, garcia_tracking_2007}. 

\par Nevertheless, detecting the acoustic modes is key to knowing more about the stars. Acoustic modes are excited by turbulent convection, which injects power into the oscillations. As all solar-like stars display an external convective zone, we expect to observe acoustic modes in all those stars. However, magnetic activity has proven to hinder the detection of the modes in \textit{Kepler} data \citep{chaplin_evidence_2011, mathur_revisiting_2019-2}. Both authors witnessed the total suppression of \textit{p}-modes in a vast part of solar-like pulsators. \cite{mathur_revisiting_2019-2} effectively detected acoustic modes in 60\% of the sample's stars. Both high magnetic field and high rotation seem to be at the origin of such modes suppression. 

\par Furthermore, amplitudes of acoustic modes are sensitive to the variations of magnetic activity: some amplitude height variations correlated with magnetic activity were first discovered in HD 49933 \citep{garcia_corot_2010}. In some other solar-like stars, there was evidence that the higher the magnetic activity, the lower the detected modes heights \citep{kiefer_stellar_2017, santos_signatures_2018, salabert_frequency_2018}. The interplay between convection and magnetic fields is currently modelled through and "on-off" approach: when the magnetic field is too strong, it completely suppresses convection. \citep[see e.g.][]{gough_influence_1966, macdonald_magnetic_2019, jermyn_origin_2020}. Integrating magnetism in theoretical models is thus essential to allow a better understanding of the oscillations and their detectability for the forthcoming space missions such as \textit{PLATO} \citep{rauer_plato_2014-2, goupil_predicted_2024}.

\par In all stars, the oscillations' amplitudes result from a balance between damping and driving. In solar-like stars, stochastic excitation by turbulence inside the convection zone is the dominant driving mechanism. Turbulent motion of stellar fluid and fluctuations of thermodynamical quantities generate power, which in turn is injected into resonant oscillation modes in the stellar cavity. \cite{lighthill_sound_1952} and \cite{stein_heating_1967} first demonstrated how a turbulent background could generate acoustic waves. \cite{goldreich_solar_1977}, \cite{balmforth_solar_1992-1}, \cite{samadi_excitation_2001}, \cite{belkacem_stochastic_2008}, \cite{philidet_modelling_2020} formalised stochastic excitations for p-modes in stars linearising the equations of motion in a turbulent background. \cite{press_radiative_1981}, \cite{goldreich_wave_1990}, \cite{belkacem_mode_2009} and then \cite{lecoanet_internal_2013} elaborated a model for gravity modes (g-modes) under a similar framework. All these methods differ from each other in the simplifications and approximations they adopt, as well as the way turbulent convection is described. Damping, which is due to several complex non-adiabatic and turbulent processes, has been studied widely \citep{goldreich_thermal_1991, balmforth_solar_1992,grigahcene_convection-pulsation_2005, belkacem_damping_2012}. All these formalisms ignore both rotation and magnetism.
\par All these studies make use of modelisation of turbulent convection, with different approximations \citep{frisch_turbulence_1995, tennekes_first_1972, lesieur_turbulence_2008}. Homogeneous and isotropic turbulence, with several models for turbulent spectra, are commonly used. Characteristic lengths and velocities in the turbulent flow can be assessed with Mixing Length Theory (hereafter MLT), by considering these values are imposed by the eddies that carry the most heat \citep{bohm-vitense_uber_1958, gough_mixing-length_1977}. 
\par However, convection is strongly affected by both rotation and magnetic fields \citep{barker_theory_2014, hotta_breaking_2018}. Convection in magnetized and rotating stars has been studied through global numerical simulations \citep[e.g.][]{brun_globalscale_2004, kapyla_local_2005, brown_rapidly_2008, charbonneau_dynamo_2010, augustson_convection_2012}. From a theoretical point of view, \cite{stevenson_turbulent_1979} adapted the Mixing Length Theory to give a prescription for the modification of the characteristic convective velocity and wavenumber in rotating and magnetized convection. \cite{barker_theory_2014, currie_convection_2020} have found good agreement between \cite{stevenson_turbulent_1979} prescriptions and numerical simulations of rotating convection. Furthermore, \cite{augustson_model_2019} generalised the theory of \cite{stevenson_turbulent_1979} by including viscous and heat diffusions to provide with a prescription for convective penetration, which has been successfully compared with numerical simulations by \cite{korre_dynamics_2021}.
\par In addition, magnetic fields and rotation influence waves propagation, and their coupling with excitation sources. The modes frequencies and displacement are affected, both for acoustic modes \citep{belkacem_mode_2009} and gravity modes \citep{mathis_transport_2009, augustson_model_2020}. In the geophysics field, the only work that has studied the impact of magnetic fields on the stochastic excitation of oscillation has focused on Magneto-Archimede-Coriolis waves in the Earth core \citep{buffett_stochastic_2018}.

\par Revisiting the existing formalism to take into account rotation and magnetic fields is then necessary. In this first work, we neglect rotation and focus on the impact of magnetic field alone. We generalise the work by \cite{samadi_excitation_2001} taking the magnetic field into account. First, in section \ref{sec:stocha}, we present the general set-up that describes stochastic excitation for both \textit{g} and \textit{p}-modes, in magnetised stars. In particular, we display the additional source terms that inject power into the modes. We assess the impact of these terms using scaling laws in section \ref{sec:scaling_laws}. We use the outcomes of \cite{stevenson_turbulent_1979} to account for the modification of turbulent convection by a magnetic field, and compare it critical magnetic field approach. (section \ref{sec:magnetized_convection}). 
Finally, in section \ref{sec:magnetized_convection_amplitude}, we assess the impact of magnetised convection on the driving and damping of oscillations. We discuss the consequences for asteroseismology, as well as the next theoretical and observational work to come (section \ref{conclusion}).

\section{Turbulent stochastic excitation in presence of magnetic field}
\label{sec:stocha}
\subsection{The inhomogeneous wave equation}
\label{sub:inhomogeneous}
\subsubsection{System of equations}
Following the method of \cite{samadi_excitation_2001}, we derive the inhomogeneous wave equation by taking into account the Lorentz force. We use the usual spherical coordinates $(r,\theta,\varphi)$ and the corresponding unit vector basis ($\er, \et, \ep$). \\
Here, $\boldsymbol{u}$ is the velocity field associated with both turbulent convective and oscillations motions.

\par The mass conservation equation writes: 
\begin{equation}
 \Dt{\rho} + \di{(\rho \boldsymbol{u})} = 0, 
 \label{continuity}
\end{equation}
where  $\rho$ is the plasma density.
The equation of momentum, while taking into account the Lorentz force boils down to: 

\begin{equation}
    \Dt{(\rho \boldsymbol{u})} + \nab: (\rho \boldsymbol{u}\boldsymbol{u})  = \rho \boldsymbol{g} - \nab P + \frac{(\nab \times \B) \times \B}{\mu_0},
     \label{momentum}
\end{equation}
where $P$ the pressure, $\boldsymbol{g}$ the gravitationnal field, and $\boldsymbol{B}$ the magnetic field. As a first step, we neglect the effects of rotation.

\par One finally needs the magnetohydrodynamics induction equation to account for the coupling between velocity and magnetic field: 

\begin{equation}
    \Dt{B} = \nab \times (\boldsymbol{u}\times \B) + \eta_B \Delta \B,
    \label{eq:induction}
\end{equation}
where $\eta_B$ is the Ohmic diffusivity, which we assume here to be uniform to simplify the writing.

\subsubsection{Physical hypothesis and decomposition}

\par To study the source term due to the turbulent background, we split each physical quantity $f$ into its equilibrium value $f_0$, and an Eulerian fluctuation $f_1$. Using the Cowling approximation \cite{cowling_non-radial_1941} ($\boldsymbol{g_1} = 0$), and assuming that equilibrium quantities verify the hydrostatic equilibrium equation, we obtain the equations for the perturbations :

\begin{equation}
\Dt{\rho_1} + \di [{(\rho_0 + \rho_1) \boldsymbol{u}}] = 0,
\label{continuity_perturbed}
\end{equation}

\begin{dmath}
    \Dt{((\rho_0 + \rho_1) \boldsymbol{u})} + \nab: ((\rho_0 + \rho_1) \boldsymbol{u}\boldsymbol{u}) = \rho_1 \boldsymbol{g_0} - \nab P_1 + \frac{(\nab \times \B) \times \B}{\mu_0}.
     \label{momentum_perturbed}
\end{dmath}

The perturbed equation of state, using an order 2 expansion, writes: 
\begin{equation}
    P_1 = c_s^2 \rho_1 + \alpha_s s_1 + \alpha_{\rho \rho}\rho_1^2 + \alpha_{ss}s_1^2 + \alpha_{\rho s} \rho_1 s_1 , 
    \label{state_equation}
\end{equation}
where: 
$$
\begin{aligned}
\alpha_s & =\left(\frac{\partial P}{\partial s}\right)_\rho, & \alpha_{\rho s}=\left(\frac{\partial^2 P}{\partial \rho \partial s}\right), \\
\alpha_{s s} & =\left(\frac{\partial^2 P}{\partial s^2}\right)_\rho, &\alpha_{\rho \rho}=\left(\frac{\partial^2 P}{\partial \rho^2}\right)_s, \\
c_s^2 & =\Gamma_1 P_0 / \rho_0.
\end{aligned}
$$
We introduce $c_s$ the sound waves velocity and $\Gamma_1= \left(\frac{\partial \ln P }{\partial \ln \rho}\right)_s$ is the first adiabatic exponent, with $s$ the macroscopic entropy. We assume that the oscillations follow an adiabatic evolution so that the Lagrangian entropy fluctuation is exclusively due to turbulence. We denote $\delta s_1$ the Lagrangian entropy fluctuations, and $s_1$ the Eulerian fluctuations. The link between the Eulerian and Lagrangian entropy fluctuations is: 

\begin{equation}
    \frac{d \delta s_t}{dt} = \Dt{s_1} + \adv{u}{(s_0+s_1)}.
    \label{lagrangian_eulerian_entropy}
\end{equation}

Combining the time-derivative of Eq.(\ref{state_equation}) with Eq. (\ref{lagrangian_eulerian_entropy}), one finds: 

\begin{dmath}
    \Dt{P_1} = \alpha_s \frac{d \delta s_1}{dt} - \alpha_s \adv{u}{(s_0 + s_1)} + (c_s^2+ 2 \alpha_{\rho \rho} \rho_1 + \alpha_{\rho s} s_1) \Dt{\rho_1} + (2 \alpha_s s_1 + \alpha_{\rho s} \rho_1) \Dt{s_1}.
    \label{state_entropy}
\end{dmath}

\par To assess the impact of turbulence on excited modes, we consider that the velocity field is composed of the oscillation velocity  $\boldsymbol{u}_{\rm osc}$ and the convective turbulent velocity $\boldsymbol{U}_t$ such that: 
\begin{equation}\boldsymbol{u} = \boldsymbol{U}_t + \boldsymbol{u}_{\rm osc}.\end{equation}
The magnetic field has three components: a large-scale magnetic field $\bar{\boldsymbol{B}}$, a small-scale turbulent magnetic field $\boldsymbol{B_{t}}$, and a fluctuation of the magnetic field induced by the oscillation motion $\boldsymbol{b_{\rm osc}}$: 

\begin{equation}
\boldsymbol{B} = \bar{\boldsymbol{B}} + \boldsymbol{B_t} + \boldsymbol{b_{\rm osc}}.
\end{equation}

It is important to note the difference between $\bar{\boldsymbol{B}}$ and $\boldsymbol{B_t}$: $\bar{\boldsymbol{B}}$ is the large-scale fields that vary over long timescale when compared to the oscillation period. It modifies the hydrostatic equilibrium inside the star, and thus the oscillations resonant cavity \citep[e.g.][]{duez_effect_2010}. 
$\boldsymbol{B_t}$ is the fluctuating part of the dynamo field and acts on small lengths and short time scales. It acts as a source term for the stochastic excitation of waves, as demonstrated in Geophysics \citep[e.g.][]{buffett_stochastic_2018}. In Eq. (\ref{eq:induction}), the last diffusive term characterises a the decrease of the magnetic field because of the Ohmic diffusion. In a Sun-like star, the Ohmic diffusion timescale associated with the acoustic oscillations is  $\tau_{\rm ohm, osc} = \ell_{\rm osc}^2/\eta_B \sim 10^{7}s$, where $\ell_{\rm osc}$ is the typical lengthscale for the oscillation computed with GYRE oscillation code \citep{townsend_gyre_2013}. The oscillation period is of the order of $\tau_{\rm osc} \sim 10^{3}s$. We then neglect the Ohmic diffusion acting on the fluctuation of the magnetic field associated with the stellar oscillations, which acts on long timescales compared to the oscillation timescale. In this framework, Eq. (\ref{eq:induction}) simplifies into:
 
\begin{align}
     \frac{\partial \bigg( \boldsymbol{B}_0 + \boldsymbol{B}_t + \boldsymbol{b}_{\rm osc} \bigg)}{\partial t} &= \nabla \times \bigg((\boldsymbol{u}_{\rm osc} + \boldsymbol{U}_t) \times (\boldsymbol{B}_0 + \boldsymbol{B}_t + \boldsymbol{b}_{\rm osc}) \bigg) \notag \\
     &\quad + \eta_B \Delta^2 (\boldsymbol{B}_0 + \boldsymbol{B}_t).
\end{align}

\par To model the stochastic excitation mechanism of acoustic waves, we need to isolate the terms due to the oscillations from the ones related to the turbulent medium in order to obtain a forced wave equation. As in the previous works \citep[e.g.][]{samadi_excitation_2001}, this leads us to make several simplifying assumptions. First, we neglect non-linear terms in oscillating quantities to recover the usual linear wave equation for acoustic modes without forcing. As a consequence, the Eulerian fluctuations of entropy and density can be seen as due only to turbulence when it comes to the forcing terms and damping terms. We then substitute $\rho_1$ (resp. $s_1$) by $\rho_t$ (resp. $s_t$) in these terms. Second, we suppose that the turbulent medium evolves freely and is not perturbed by the oscillations. We then assume two separate continuity equations, applied to turbulent and oscillating quantities independently:
\begin{equation}
    \Dt{\rho_t} = -\di \left((\rho_0 + \rho_t) \boldsymbol{U}_t \right),
    \label{continuity_turbulent}
\end{equation}
and: 
\begin{equation}
    \Dt{\rho_{osc}} = -\di \left(\rho_0\ \boldsymbol{u}_{\rm osc}\right).
    \label{continuity_osc}
\end{equation}

\noindent Finally, we use the framework of incompressible turbulence ($\di \boldsymbol{U}_t = 0$) to further simplify the modelling of the stochastic excitation.
\par Differentiating Eq. (\ref{momentum_perturbed}) with respect to time and making use of equations (\ref{continuity_perturbed}, \ref{eq:induction}, \ref{state_entropy}) while neglecting the non-linear terms in oscillating quantities yields the forced oscillation equation, also called the inhomogeneous wave equation :
\begin{equation} \left(\frac{\partial^{2}}{\partial t^{2}}-\mathcal{L} \right)(\boldsymbol{u}_{\rm osc})+\mathcal{D} (\boldsymbol{u}_{\rm osc}, \boldsymbol{U}_t)=\frac{\partial \mathcal{S}(\boldsymbol{U}_t, \boldsymbol{B_t})}{\partial t} + \frac{\partial \mathcal{C}(\boldsymbol{U}_t, \boldsymbol{B_t})}{\partial t}.
\label{eq:inhomogeneous}
\end{equation}

\par Here, $\mathcal{L}$ is the linear operator which governs the propagation of stellar oscillations, $\mathcal{D}$ is the damping term, $\mathcal{S}$ is the excitation source term, and $\mathcal{C}$ contains the source terms that are negligible, or do not contribute to the excitation. 

\par
\begin{figure}[]
    \centering
    \includegraphics[width=0.5\textwidth, trim={2cm 3cm 5cm 2cm},clip]{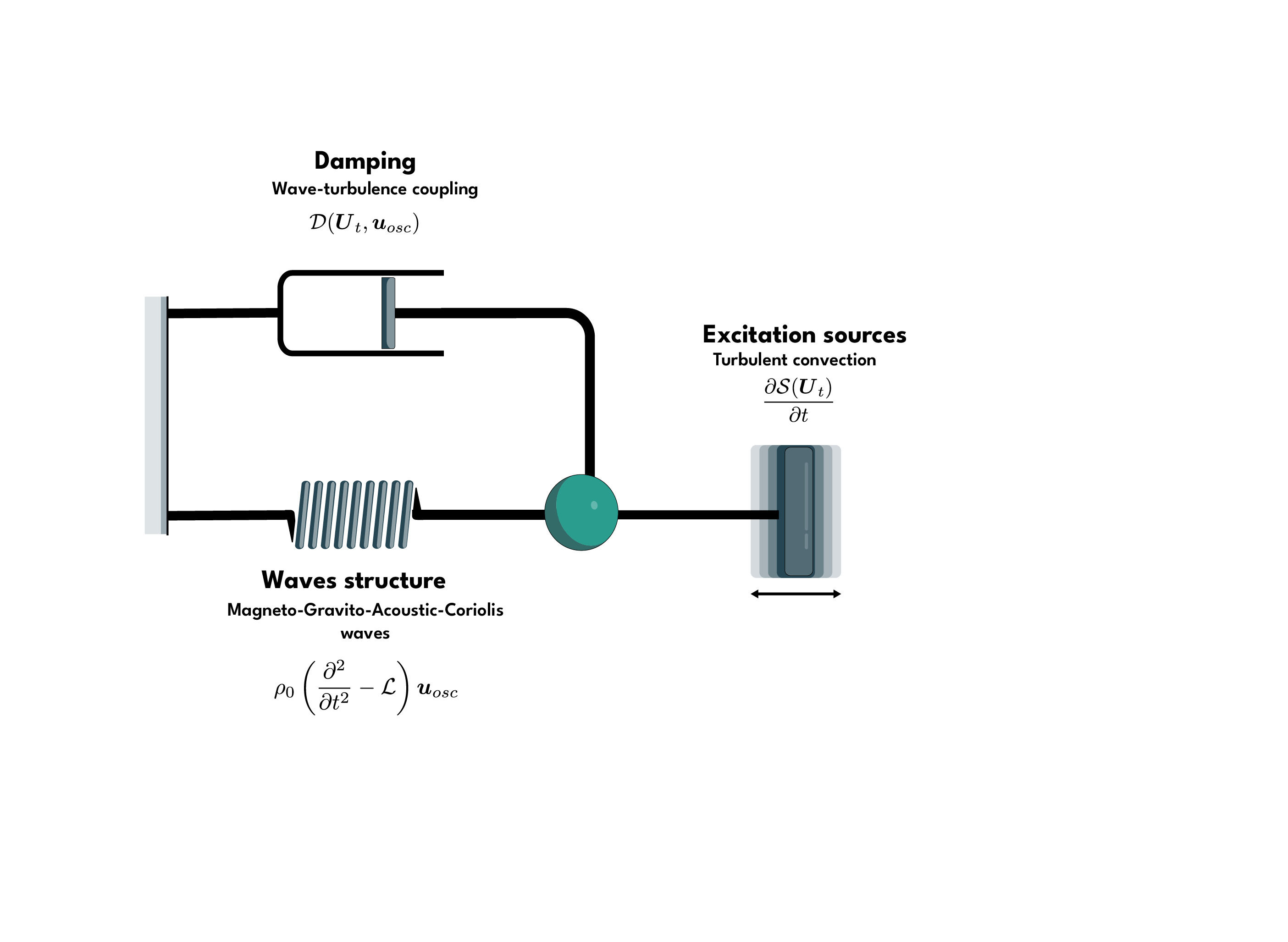}
    \caption{Analogy between the forced and damped mass-spring system, and the inhomogeneous wave equation}
    \label{fig:mase-ressort}
\end{figure}

 There is a strong physical analogy between this inhomogeneous equation and the mechanical forced mass-spring system (see fig. \ref{fig:mase-ressort}): $\mathcal{D}$ is the homologous of a mechanical damping, while $\mathcal{S}$ is the excitation source which forces the oscillations.

\subsubsection{Source terms}

\par The source term $\mathcal{S}$ accounts for the forcing of waves by turbulent fields. We can break it into several components to identify the various origins of the energy injection into the modes: 
\begin{equation}
    \mathcal{S} = \mathcal{S}_R + \mathcal{S}_S + \mathcal{S}_B,
\label{sources}
\end{equation}
where $\mathcal{S}_R$ is the Reynolds-stresses source term: 
    \begin{equation}
    \mathcal{S}_R= - \nabla: \Big( \rho \boldsymbol{U}_t \boldsymbol{U}_t \Big),
    \label{eq:reynolds_stress}
    \end{equation}
and $\mathcal{S}_S$ is an entropy fluctuations source term :
     \begin{dmath}
    \Dt{\mathcal{S}_S} = - \di ( \rho \boldsymbol{U}_t)\boldsymbol{g}_0 
    - \boldsymbol{\nabla} \left\{ \alpha_s  \frac{d \delta s_t}{dt} -  \alpha_s \adv{\boldsymbol{U}_t}{s_t} - \alpha_s \adv{\boldsymbol{U}_t}{s_0} \\
    -\di (\rho \boldsymbol{U}_t) \left[c_s^2 + 2 \alpha_{\rho \rho} \rho_t + \alpha_{\rho s} s_t \right] + 
    \left( 2 \alpha_{ss} s_t + \alpha_{\rho s} \rho_t \right) \Dt{s_t} \right \}.
    \end{dmath}
    It has been shown in recent works \citep{samadi_stellar_2015} that this entropy source accounts for less than 10 $\%$ of the total power injected into the modes. It is then negligible in solar-like stars in a first step. \\
    $\mathcal{S}_B$ is the source term due to the magnetic field, i.e. the turbulent Maxwell-stresses: 
    \begin{equation}
        \mathcal{S}_B =  \nabla: \left( \frac{B_{t,i} B_{t,j}}{\mu_0} - \delta_{ij}\frac{B_{t}^2}{2 \mu_0} \right),
        \label{eq:sourcemaxwell}
    \end{equation}
    where $\delta_{ij}$ is the Kronecker symbol, and we use Einstein summation convention for the tensorial form of the stresses. 

\subsection{Linear wave operator}
\par The linear wave operator $\mathcal{L}$ is decomposed as :

\begin{equation}
    \mathcal{L} = \mathcal{L}_p + \mathcal{L}_g + \mathcal{L}_b,
\end{equation}
where $\mathcal{L}_p$ corresponds to the acoustic character of the waves :
\begin{equation}
    \mathcal{L}_p(\boldsymbol{u}_{\rm osc}) =  -\boldsymbol{g}_{0} \boldsymbol{\nabla} \cdot\left(\rho_{0} \boldsymbol{u}_{\rm osc}\right) + \boldsymbol{\nabla}\left(c_{s}^{2} \boldsymbol{\nabla} \cdot\left(\rho_{0} \boldsymbol{u}_{\rm osc}\right)\right),
\end{equation}
$\mathcal{L}_g$ accounts for the gravity waves behavior :
\begin{equation}
    \mathcal{L}_g (\boldsymbol{u}_{\rm osc}) = \boldsymbol{\nabla}\left(\alpha_{s} \boldsymbol{\boldsymbol{u}_{\rm osc}} \cdot \boldsymbol{\nabla} s_{0}\right),
\end{equation}
and $\mathcal{L}_b$ is the Alfvén part of the waves: 
\begin{dmath}
    \mathcal{L}_B (\boldsymbol{u}_{\rm osc}, \boldsymbol{b_{\rm osc}}) = \frac{1}{\mu_0}\Bigg[ \bigg(\di{\boldsymbol{\boldsymbol{u}_{\rm osc}}} \bigg) \bar{\boldsymbol{B}} \times \bigg(\rot{\bar{\boldsymbol{B}}} \bigg) -\adv{\bar{\boldsymbol{B}}}{\boldsymbol{u}_{\rm osc}} \times \bigg(\rot{\bar{\boldsymbol{B}}} \bigg) + \adv{\boldsymbol{u}_{\rm osc}}{\bar{\boldsymbol{B}}} \times \bigg( \rot{\bar{\boldsymbol{B}}} \bigg) \\ 
    + \bar{\boldsymbol{B}} \times \rot{\bigg( (\di{\boldsymbol{u}_{\rm osc}}) \bar{\boldsymbol{B}} \bigg)} - \bar{\boldsymbol{B}} \times \rot{\bigg( \adv{\bar{\boldsymbol{B}}}{\boldsymbol{u}_{\rm osc}} \bigg)}  + \bar{\boldsymbol{B}} \times \rot{\bigg(\adv{\boldsymbol{u}_{\rm osc}}{\bar{\boldsymbol{B}}} \bigg)} - \eta_B \boldsymbol{b_{\rm osc}} \times \rot{\bigg( \Delta \bar{\boldsymbol{B}} \bigg)} \Bigg].
\end{dmath} 
Note that the oscillation velocity $\boldsymbol{u}_{\rm osc}$ is coupled to the oscillation magnetic field $\boldsymbol{b_{\rm osc}}$, via the linearised induction equation.

\subsubsection{Damping}
The damping term can be written as: 
\begin{equation}
    \mathcal{D} = \mathcal{D}_0  + \mathcal{D}_B.
\end{equation}
 The term $\mathcal{D}_0$ is due to the non-magnetic part of the inhomogeneous wave equation, which can be recovered as in previous works such as \cite{samadi_etude_2012}, \cite{belkacem_stochastic_2008}:
    
\begin{dmath}
     \mathcal{D}_0(\boldsymbol{u}_{\rm osc}, \boldsymbol{U}_t) = \frac{\partial ^2}{\partial t^2} (\rho_t \boldsymbol{u}_{\rm osc}) - \mathcal{L}_0 (\rho_t \boldsymbol{u}_{\rm osc}) + 2 \frac{\partial}{\partial t} \left(\nabla \text{: } \rho \boldsymbol{u}_{\rm osc}\boldsymbol{U}_t \right)\\
     + \nabla \bigg[ -(2 \alpha_{\rho \rho} \rho_t + \alpha_{\rho s}s_t ) \left( \nabla \cdot (\rho_t \boldsymbol{u}_{\rm osc}) \right) \\
     - \nabla \cdot (\alpha_s s_t \boldsymbol{u}_{\rm osc}) + \alpha_s (\boldsymbol{u}_{\rm osc}\cdot \nabla) s_t + s_t (\boldsymbol{u}_{\rm osc}\cdot \nabla) \alpha_s \bigg],
 \end{dmath}
where $\mathcal{L}_0 = \mathcal{L}_p + \mathcal{L}_g$.
$\mathcal{D}_B$ is the magnetic contribution to the damping :
\begin{equation}
    \mathcal{D}_B (\boldsymbol{b_{\rm osc}}, \boldsymbol{B_t} ) = \frac{\partial}{\partial t} \left( \frac{( \rot{\boldsymbol{B_t}}) \times \boldsymbol{b_{\rm osc}}}{\mu_0} + \frac{ (\rot{\boldsymbol{b_{\rm osc}}}) \times \boldsymbol{B_t}}{\mu_0} \right).
\end{equation}

\subsubsection{Negligible source terms}

The term $\frac{\partial \mathcal{C}}{\partial t}$ contains all source terms that are negligible: 

\begin{dmath}
    \frac{\partial \mathcal{C}}{\partial t} = \frac{( \rot{\bar{B}}) \times \boldsymbol{B_t}}{\mu_0} + \frac{( \rot{\boldsymbol{B_t}}) \times \bar{B}}{\mu_0} + \mathcal{L} (\rho \boldsymbol{U}_t) - \frac{\partial^2}{\partial t} (\rho \boldsymbol{U}_t)  - \frac{\partial}{\partial t} \left( \nabla \text{: } \rho_t \boldsymbol{U}_t \boldsymbol{U}_t \right )
    + \\
    \nabla \left[ -(2 \alpha_{\rho \rho} + \alpha_{\rho s} s_t) \di (\rho \boldsymbol{U}_t) - (2 \alpha_{ss}s_t + 2 \alpha_{\rho s} \rho_t) \frac{\partial s_t}{\partial t} 
    - \alpha_s \boldsymbol{U}_t \cdot \nabla s_t \right].
\end{dmath}
It contains linear source terms in turbulent quantities, which do not contribute to the excitation, as well as terms that are negligible due to their order of magnitude \citep[see e.g.][for more details]{samadi_excitation_2001}. 

\subsection{Modes amplitudes}

\par We seek the mean square amplitude of $\boldsymbol{u}_{\rm osc}$, for each mode, of given radial, angular and azimuthal order $(n,\ell,m$ respectively). The wave Eulerian displacement is written in complex notations as the product between an instantaneous amplitude $A(t)$, and a Lagrangian eigendisplacement $\boldsymbol{\xi}(\boldsymbol{r})$, which depends only on the position \citep{samadi_excitation_2001, belkacem_mode_2009}: 
\begin{equation}
\delta \boldsymbol{r}_{osc}(\boldsymbol{r},t) = \frac{1}{2}  A(t) \boldsymbol{\xi} (\boldsymbol{r})e^{i \omega_0 t} + c.c.
\label{eq:eigen_displacement}
\end{equation}
We introduce $\omega_0$ the eigenmode frequency, and \textit{c. c.} denotes the complex conjugate. Deriving Eq. (\ref{eq:eigen_displacement}) with respect to time leads to the wave velocity: 
\begin{equation}
\boldsymbol{u}_{\rm osc}(\boldsymbol{r},t) = A(t)i \omega_0 \boldsymbol{\xi}(\boldsymbol{r}, t) e^{i \omega_0 t},
\label{eq:uosc}
\end{equation}
where we consider that $A(t)$ is evolving at a longer timescale than the wave oscillation period. Using Eqs. (\ref{eq:inhomogeneous}) and (\ref{eq:uosc}), one derives the mean squared amplitude of the oscillations:
\begin{equation}
\left\langle|A|^2\right\rangle=\frac{1}{8 \eta_D\left(\omega_0 I\right)^2} \int d^3 x_0 \int_{-\infty}^{+\infty} d^3 r d \tau e^{-i \omega_0 \tau}\left\langle(\vec{\xi} \cdot \mathcal{S})_1(\vec{\xi} \cdot \mathcal{S})_2\right\rangle.
\label{eq:amplitude_source}
\end{equation}
Subscripts 1 and 2 make reference to spatio-temporal positions $\left(\vec{x}_0-\frac{\vec{r}}{2},-\frac{\tau}{2}\right)$ and $\left(\vec{x}_0+\frac{\vec{r}}{2}, \frac{\tau}{2}\right)$ respectively. 
$\langle.\rangle$ denotes a statistical average performed on an infinite number of independent realisations. We introduce $\eta_D$ as damping, and $I$ the mode inertia: 
\begin{equation}
    I = \int_0^M \boldsymbol{\xi} \cdot \boldsymbol{\xi}^\star dm, 
\end{equation}
where $dm$ is the mass of an elementary fluid parcel.
Furthermore, we assume turbulence is stationary and homogeneous so that the source term $\mathcal{S}$ is invariant by any time translation. Although turbulence in magnetohydrodynamics is known to be anisotropic, this assumption is justified by the fact that most of the excitation is due to small-scale eddies \citep[see e.g.][]{samadi_excitation_2001}. 

 \par There are several contributions to the excitation power: $\mathcal{S}_R^2$ and $\mathcal{S}_B^2$ correspond to the power injected by the Reynolds and Maxwell stresses, respectively. A cross-term also emerges with the combinations in  $\mathcal{S}_B \mathcal{S}_R$. 

\par
\begin{figure}[h]
    \centering
\includegraphics[width=0.49\textwidth]{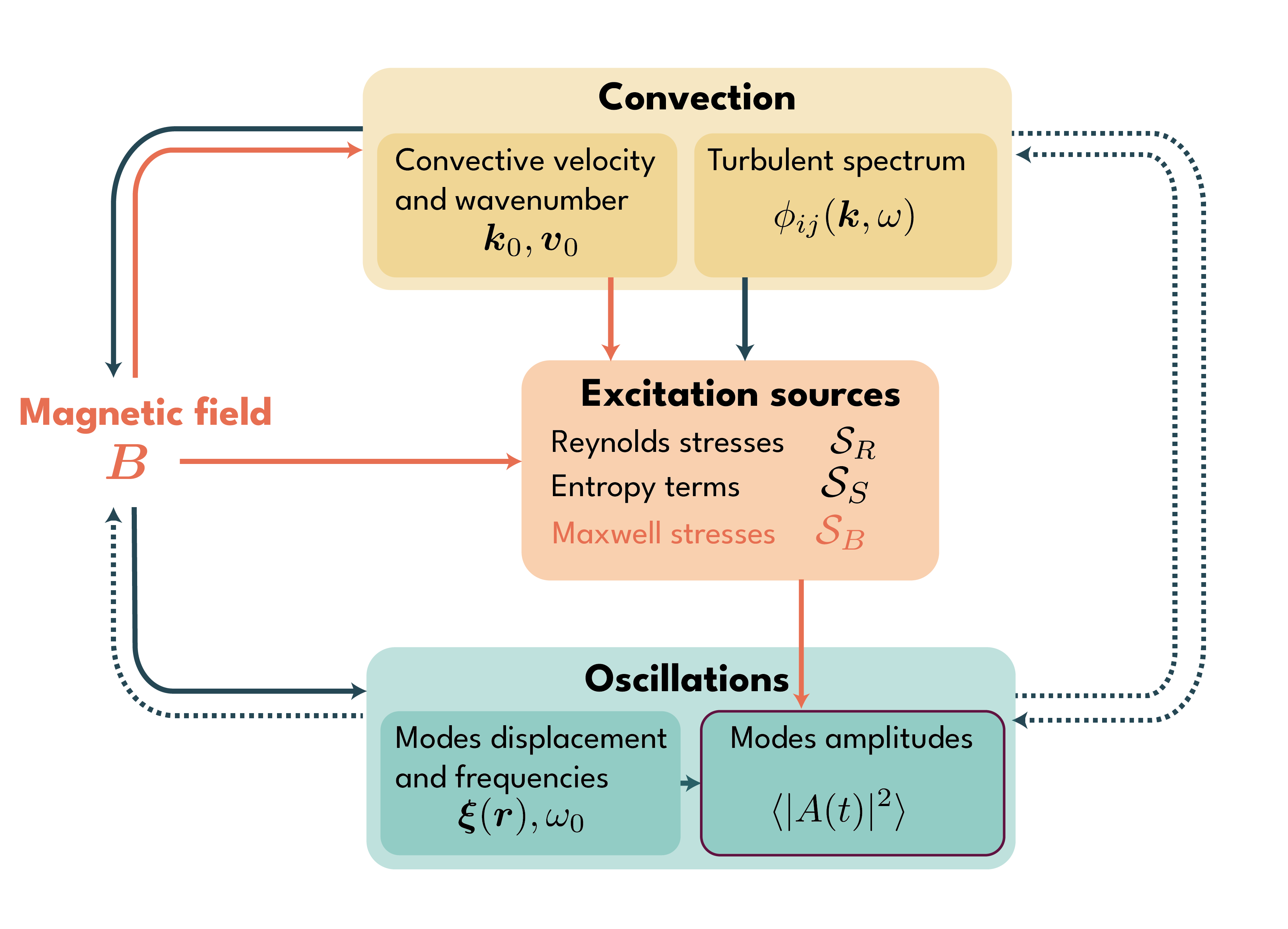}
    \caption{Influence of the magnetic field on modes amplitudes: different phenomena at stake. The red arrows represent the effect we take into account in the present paper. }
    \label{fig:schema-mag}
\end{figure}

 Figure \ref{fig:schema-mag} illustrates the complex interdependencies between magnetic field and modes'amplitudes: not only does the magnetic field add up a source term compared with the non-magnetised case, but it also influences indirectly the existing source terms: 
\begin{enumerate}
    \item The stellar structure is modified by the magnetic force \citep[see e.g.][]{duez_effect_2010}. The star's density and thermodynamic quantities profiles, and in turn, the mode frequencies change. This is the so-called indirect effect on stellar oscillations \citep{gough_effect_1990}.
    \item Waves are affected by the magnetic field so that they become magneto-acoustic-gravity waves. This changes both the frequencies of the modes $\omega_0$ which are shifted, and the eigenfunctions of the displacement $\boldsymbol{\xi}$. This is the direct magnetic effect on stellar oscillations \citep{gough_effect_1990}.
    \item The characteristics of convection strongly affect the source terms. This happens through a change in the mean parameters from MLT such as the convective wavenumber $\boldsymbol{k}_0$ and the convective velocity $\boldsymbol{v}_0$. Magnetic field tends to modify convection, by changing the instability threshold \citep{chandrasekhar_hydrodynamic_1961}. The magnetic field also affects turbulent convection through a change in the kinetic energy spectrum and magnetic energy spectrum. \citep[e.g.][]{brun_interaction_2004}. This effect of magnetic field on convection is paramount in the present model, and will be widely discussed in Section \ref{sec:magnetized_convection}.
\end{enumerate}
This complex interplay between magnetic field and excitation has not been studied in the literature. To give a first overview of the impact of magnetic fields on the stochastic excitation of modes, we will as a first step make use of scaling laws for modulation of the modes amplitudes as a function of the strength of the magnetic field. Such scalings prescriptions could be directly compared to observations. We first adopt a Mixing Length Theory approach to model convection.

\section{Scaling laws for the power injected into the modes}
\label{sec:scaling_laws}
To give a very first overview of the impact of magnetism on the stochastic excitation of waves,  we make use of scaling laws to evaluate the modes amplitudes. Indeed, no physical model up to this date includes the effects of magnetic fields. Scaling relationships have proven to give relevant tendencies and are often used in the litterature, for example in dynamo theories \citep[e.g.][]{augustson_rossby_2019}. 
When it comes to p-modes, the energy provided by the stochastic excitation is mostly injected very locally at a radius $\boldsymbol{r}_i$, in the photosphere \citep[e.g.][]{samadi_excitation_2001}. As gravity modes are evanescent in the convective zone, they are mostly excited at the base of the convective zone. Under this assumption, eq. (\ref{eq:amplitude_source}) boils down to: 
\begin{equation}
\begin{aligned}
&\left \langle|A|^2\right\rangle=\frac{1}{8 \eta_D \left(\omega_0 I\right)^2} \int d^3 x_0 \times \\
&\int_{-\infty}^{+\infty} (\boldsymbol{r} - \boldsymbol{r}_i) r  d^3 \delta d \tau e^{-i \omega_0 \tau}\left\langle(\vec{\xi} \cdot \mathcal{S})_1(\vec{\xi} \cdot \mathcal{S})_2\right\rangle, 
\end{aligned}
\label{eq:amplitude_dirac}
\end{equation}
where $\delta$ is the usual Dirac function.\\
We neglect in the present model the influence of the magnetic field on the eigenfunctions $\boldsymbol{\xi}$, which means that $\omega_0 \gg \omega_A$, where $\omega_A$ is the Alfvén pulsation. Under this hypothesis, one can use estimates of the source $\mathcal{S}$ and the damping $\eta_D$ at the location $\boldsymbol{r}_i$ to evaluate the resulting amplitudes of the acoustic modes. Indeed, for a given mode, one has: 
\begin{equation}
    \langle \lvert A \rvert \rangle^2 \propto \frac{\mathcal{S}^2}{\eta_D}.
\end{equation}

\subsection{Scaling of the source terms}
\label{sub:scaling_source}
First, we estimate the scaling relationship of the source terms. As the entropy source term $\mathcal{S}_s$ is negligible, we only consider the Reynolds-stresses $\mathcal{S}_R$, and the Maxwell-stresses $\mathcal{S}_B$ source terms. \\
We introduce $\ell_{osc}$, which is the typical length of the oscillations. By integrating by part eq. (\ref{eq:amplitude_source}), all the spatial gradients of the source terms Eqs. (\ref{eq:reynolds_stress}) and (\ref{eq:sourcemaxwell}) can be transferred to the oscillations, as detailed in \cite{samadi_excitation_2001, belkacem_mode_2009}.
Furthermore, the eddies that contribute the most to the stochastic excitation are the ones whose turnover time is near the wave frequency $\tau_{\rm turnover} \sim \omega_0$. For this reason, we consider any time derivative amounts to a multiplication by a factor $\omega_0$, and any spatial derivative amounts to a multiplication by $1 / \ell_{\rm osc}$. \\
We introduce $u_\lambda$, the velocity of a given convective eddy, of size $\lambda$. We also introduce $u_c$, the MLT convective velocity, which is the velocity of the largest eddy of the turbulent cascade, which has a size $\Lambda$. Similarly, $b_\lambda$ is the magnetic field associated with a given eddy of size $\lambda$, while $B_0$ is the magnetic field associated with the largest eddy of size $\Lambda$.
\par In this framework, the Reynolds-stresses source term scales like: 
\begin{equation}
    \mathcal{S}_R \sim  \frac{- \rho_0 u_\lambda^2}{\ell_{osc}} ; 
    \label{eq:scaling_s_r}
\end{equation}
and the Maxwell-stresses source term scales like: 
\begin{equation}
    \mathcal{S}_B \sim \frac{ b_\lambda^2}{2 \mu_0\ell_{osc}}.
\end{equation}
Furthermore, scaling relationships are often used in turbulence to link $u_\lambda$ and $u_c$. For a given slope $\alpha$ for the kinetic energy spectrum, one has \citep[see e.g. Chapter 1 of][]{tennekes_first_1972}: 
\begin{equation}
    u_\lambda \sim u_c \left(\frac{\lambda}{\Lambda}\right)^{(\alpha-1)/2}.
    \label{eq:cascade_u}
\end{equation}
The same can be applied to the magnetic cascade: 
\begin{equation}
    b_{\lambda} \sim B_0 \left(\frac{\lambda}{\Lambda}\right)^{(\alpha-1)/2}.
    \label{eq:cascade_b}
\end{equation}
In Eqs. (\ref{eq:cascade_u})-(\ref{eq:cascade_b}), we assume that the magnetic energy spectrum and the kinetic energy spectrum have the same slope $\alpha$. This is the case for an Iroshnikov-Kraichman spectrum in magnetohydrodynamic turbulence, for which the slope is $\alpha = -3/2$, both for the kinetic and the magnetic energy spectra \citep{iroshnikov_turbulence_1964, kraichnan_inertialrange_1965}. Depending on the turbulent regimes in magnetohydrodynamics, different values can be found for these spectra, as seen in experiments and direct numerical simulations. \citep[see e.g.][]{sommeria_experimental_1986, biskamp_scaling_2000, mininni_finite_2009}.
Finally, using equations (\ref{eq:scaling_s_r})-(\ref{eq:cascade_b}) when taking magnetic fields into account, the resulting source $\mathcal{S}_{\rm mag} = \mathcal{S}_B + \mathcal{S}_R $ scales like: 
\begin{equation}
    \mathcal{S}_{\rm mag}  \sim - \mathcal{S}_R \left(1 - \frac{V_A^2}{2 u_c^2} \right), 
\end{equation}
where we have introduced the Alfvén velocity: 
\begin{equation}
    V_A \equiv \sqrt{\frac{B_0^2}{\rho_0 \mu_0}}.
\end{equation} 
As highlighted in Fig. \ref{fig:schema-mag}, a magnetic field also has an indirect effect on the stochastic excitation as it influences convection. The convective velocity $u_c$ must then be evaluated in the framework of magnetised convection, and is then expected to be potentially weaker than in non-magnetised convection.

\subsection{Scaling of the damping rate}

As shown in Eq. (\ref{eq:amplitude_source}), knowing the damping coefficient $\eta_D$ is paramount when it comes to assessing the modes mean amplitudes (see Fig. \ref{fig:mase-ressort}). This coefficient is proportional to the wave's loss of energy. We consider in a first step that damping comes from two phenomena: a turbulent viscous dissipation of the wave, and its Ohmic dissipation. The convective zone of solar-like stars is known to be a turbulent medium. The molecular viscosity is then negligible in front of the turbulent eddies viscosity: $\nu \sim \nu_{\rm turb}$. However, more complex formalisms have emerged taking into account the non-adiabatic fluctuations of density, entropy and turbulent pressure inside the star \citep[e.g.][]{grigahcene_convection-pulsation_2005, belkacem_damping_2012}. To compare hydrodynamical and magnetic effects we restrict ourselves as a first step to the simplest eddy viscosity and Ohmic diffusivity modeling.

The volumic loss of energy due to a turbulent viscous dissipation for the wave is: 
    \begin{equation}
    \mathcal{D}_{vis} = \rho_0 \nu_t (\boldsymbol{\nabla} \times \boldsymbol{u}_{\rm osc})^2,
    \end{equation}
    where $\nu_t$ is the eddy viscosity. The turbulent viscosity scales like: 
    \begin{equation}
        \nu_{\rm turb} \sim \frac{\ell_c u_c}{3},
        \label{eq:nu_turb}
    \end{equation}
    where $\ell_c$ is the convective characteristic length in MLT.

The volumic loss of energy due to Ohmic dissipation for the wave is 
    \begin{equation}
    \mathcal{D}_{ohm} = \mu_0 \eta_B \boldsymbol{j}_{osc}^2,
    \end{equation}
    where $\boldsymbol{j}_{osc}$ is the current density associated with the wave, $\mu_0$ the vacuum magnetic permeability and $\eta_B$ the magnetic diffusivity.

Using the same notations and the same methodology as in Subsection \ref{sub:scaling_source}, we find: 
\begin{equation}
\mathcal{D}_{vis} \sim \frac{\rho_0 \nu_{\rm turb} \boldsymbol{u}_{\rm osc}^2}{\ell_{osc}^2}.
\end{equation}
We make use of the induction equation (\ref{eq:induction}), while neglecting the diffusive term, to relate the wave's magnetic field $\boldsymbol{b_{\rm osc}}$ and the large-scale equilibrium magnetic field $\bar{\boldsymbol{B}}$. We find :

\begin{equation}
    \boldsymbol{b}_{\rm osc} = \boldsymbol{\nabla} \times (\xi \times \bar{\boldsymbol{B}}). 
\end{equation}
Furthermore: 
\begin{equation}
    \xi \sim \frac{\boldsymbol{u}_{\rm osc}}{\omega_0}. 
\end{equation}
As $\boldsymbol{j}_{\rm osc} = \frac{\boldsymbol{\nabla} \times \boldsymbol{b}_{\rm osc}}{\mu_0}$ from Maxwell-Ampere equation, we derive the scaling relationship for the Ohmic dissipation: 
\begin{equation}
\mathcal{D}_{\rm ohm} \sim \frac{\eta_B \boldsymbol{u}_{\rm osc}^2 \bar{\boldsymbol{B}}^2}{\omega_0^2 \mu_0^2 \ell_{\rm osc}^4}.
\end{equation}
Moreover, we consider that the magnetic field at the injection scale in the turbulent cascade comes from the large-scale magnetic field, such that: 
\begin{equation}
    \boldsymbol{B}_0 \sim \bar{\boldsymbol{B}}.
\end{equation}
One can express the characteristic oscillation length with the dispersion relation for acoustic waves: 
\begin{equation}
    \ell_{osc} \sim \frac{c_s}{\omega_0},
\end{equation}
where $c_s$ is the local sound speed.
When taking magnetism into account, the resulting damping contribution is $\mathcal{D}_{\rm mag} = \mathcal{D}_{\rm ohm} + \mathcal{D}_{\rm vis}$. We have: 
\begin{equation}
    \mathcal{D}_{mag} \sim \mathcal{D}_{vis} \left(1 +  \frac{A^2 \mathcal{M}^2}{\mathcal{P}_m} \right), 
\end{equation}
where $\mathcal{P}_m \equiv \nu / \eta_B$ is the magnetic Prandtl number, and $\mathcal{M} \equiv u_c / c_s$ is the local Mach number. We introduced the dimensionless inverse Alfvén number, which corresponds to the ratio between magnetic energy and convective kinetic energy: 
\begin{equation}
    \displaystyle A = \frac{V_A^2}{u_c^2} = \frac{B_0^2}{\rho_0 \mu_0 u_c^2}.
\end{equation}
Again, the convective velocity $u_c$ depends on the strength of the magnetic field. To assess the amplitudes of the mean modes appropriately, one needs a proper modelling of magnetised convection.

\section{Magnetised convection}

\label{sec:magnetized_convection}
\subsection{Modelling magnetised convection}
\par Magnetic fields modify convection, which in turn impacts the excitation sources and damping. The impact of rotation and magnetic fields on convection is of keen interest for stellar and planetary evolution \citep[see e.g.][]{maeder_physics_2009}. The most realistic approach to modelling stars and planets
convective zone is to solve the underlying equations numerically \citep[e.g.][]{brown_magnetic_2011, brun_magnetism_2017, brun_powering_2022}. 

\par However, as the computational resources available are finite, a local model of convection is often used, particularly for stellar structure modelling and secular evolution. Pioneering studies by \cite{chandrasekhar_hydrodynamic_1961} and \cite{canuto_stellar_1991} showed that both rotation alone and magnetic fields alone tend to inhibit the convection'strength. However, rotation and magnetic field acting together tend to destabilise convection with respect to the case of a magnetised but no-rotating fluid or a rotating but non-magnetised fluid \citep{horn_elbert_2022}. The interplay between these two phenomena is complex. In this first work, we focus on the case of the sole magnetic field. 
There are two main theoretical approaches to account for the impact of magnetic fields on convection: an instability criterion approach and the Mixing-Length Theory approach.
\subsubsection{Instability criterion.} Knowing under which conditions a star is unstable to convection is key. \cite{schwarzschild_equilibrium_1906} and \cite{ledoux_stellar_1947} derived instability criteria for convection to start, which are used in stellar models. As explained before, convection is stabilised by magnetic fields. At some point, if the external magnetic field is too high, the fluid becomes stable to convection: we consider that convection is frozen in this case \citep{gough_influence_1966}. The value of the minimum magnetic field that freezes convection is the critical magnetic field $B_{\rm crit}$. This approach is an \textit{"on-off"} approach to magnetoconvection: either the external magnetic field $B$ is below $B_{\rm crit}$ and convection is active, or $B > B_{\rm crit}$ and there is no more convection (see the illustration Fig. \ref{fig:critical_magnetic}). It is often used in stellar physics to tackle strongly magnetised stars \citep[see e.g.][]{jermyn_origin_2020}.
 \subsubsection{Mixing-Length Theory} 
\begin{figure}[h]
    \centering
    \begin{subfigure}[b]{0.5\textwidth}
        \centering
        \includegraphics[width=\textwidth]{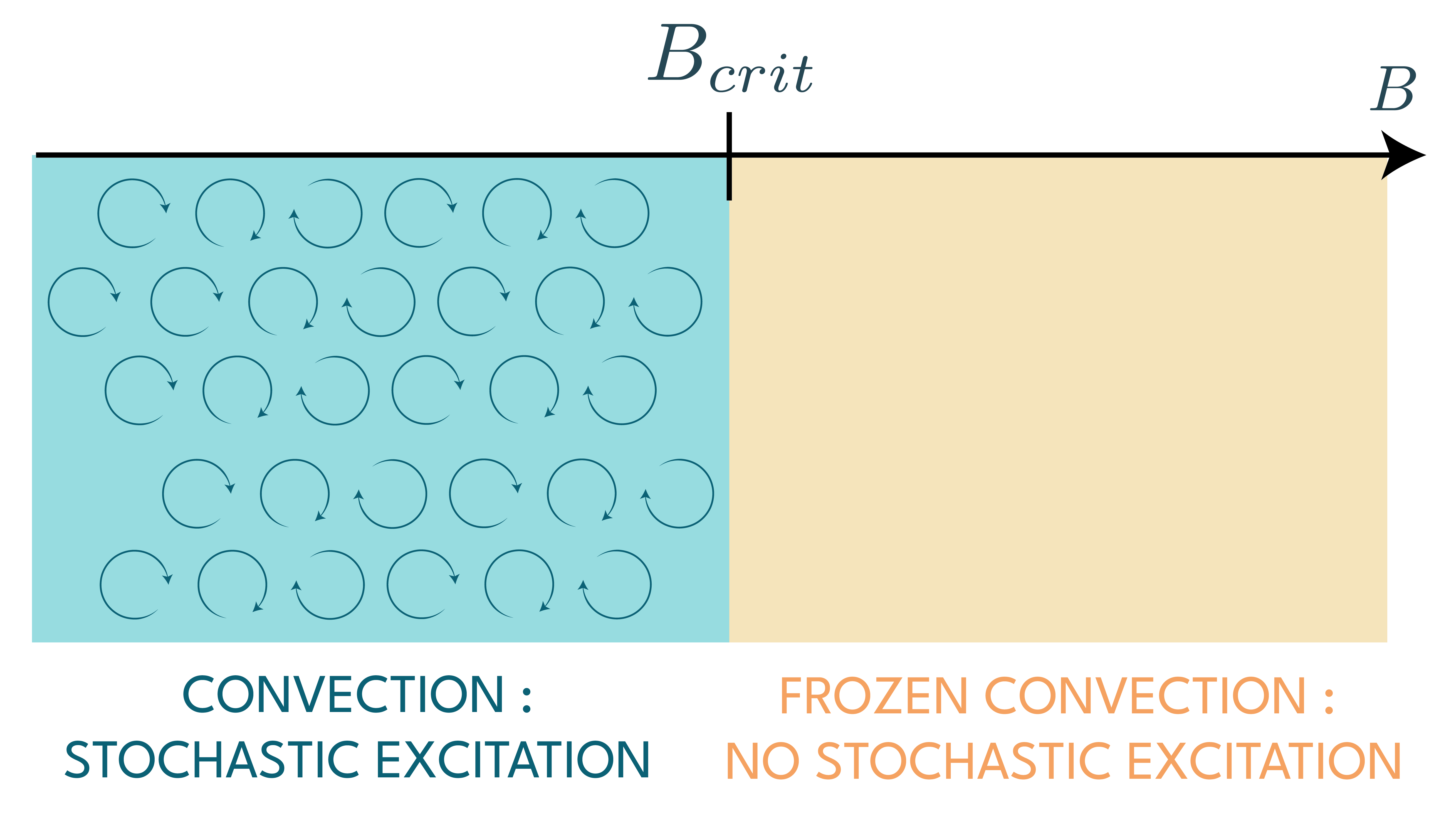}
        \caption{Critical magnetic field in magnetised convection.}
        \label{fig:critical_magnetic}
    \end{subfigure}
    
    \begin{subfigure}[b]{0.5\textwidth}
        \centering
        \includegraphics[width=\textwidth]{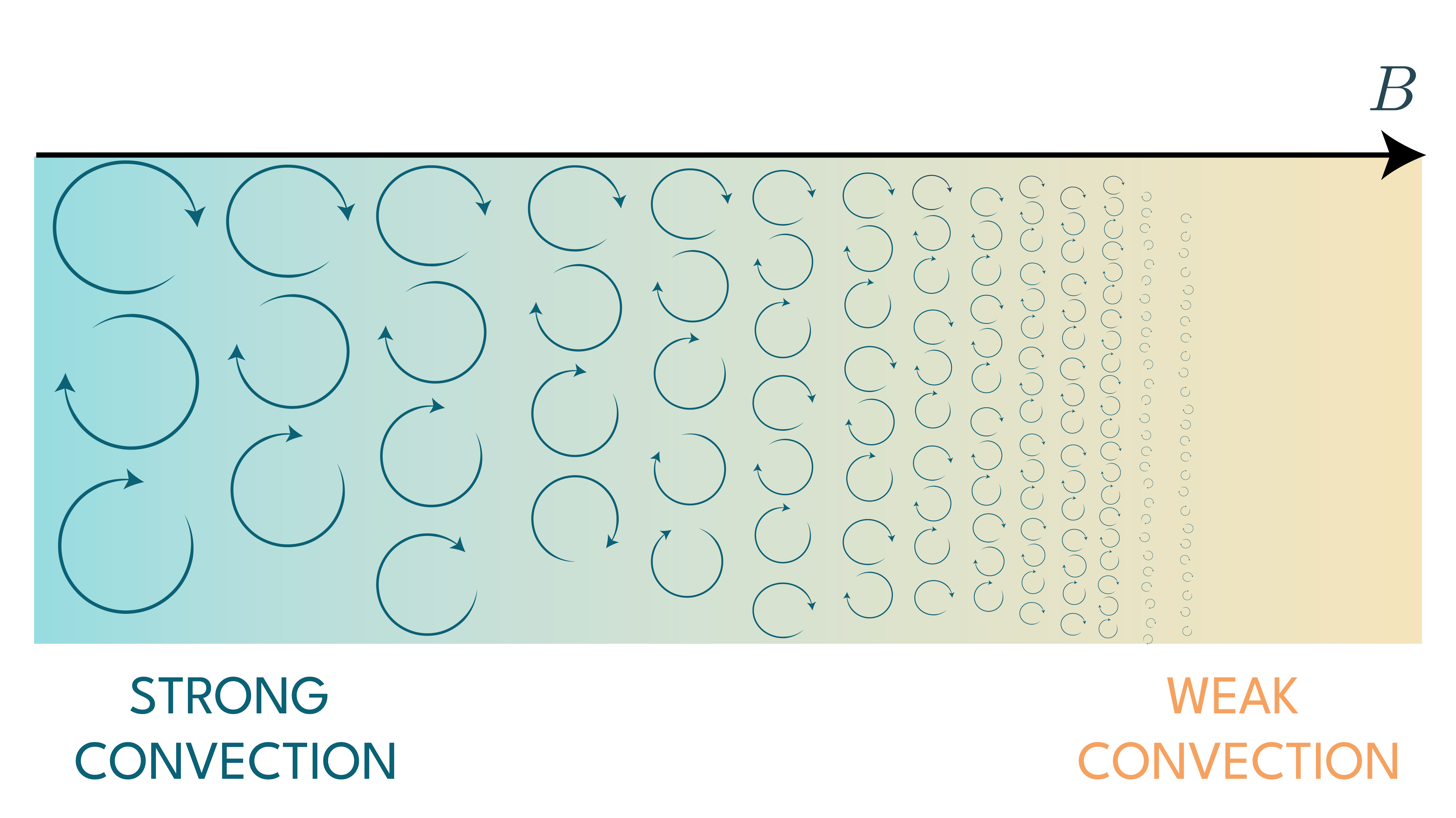}
        \caption{Magnetic Mixing Length Theory.}
        \label{fig:mmlt}
    \end{subfigure}

    \caption{Treatment of the impact of magnetic field on convection, with the critical magnetic field (a), and the Magnetic Mixing- Length Theory (b).}
    \label{fig:magnetised_convection}
\end{figure}

Modelling turbulent convection is quite complex due to a large number of space and time scales. Mixing-Length Theory (hereafter MLT) \citep{bohm-vitense_uber_1958, gough_mixing-length_1977} models convection with a single length and velocity scale, which corresponds to the characteristics of the most energetic convective eddy. Currently, MLT is implemented in 1-D stellar evolution codes to provide with the main properties of stellar convection zones.\\
 The critical magnetic field approach described in the previous section does not account for the progressive diminution of the convection'strength when increasing an external magnetic field. When increasing the magnetic field, the turbulent eddies become smaller and smaller, and their characteristic velocity diminishes. \cite{stevenson_turbulent_1979} derived a prescription accounting for the modification of Mixing-Length Theory by an external magnetic field, rendering the progressive reduction of the convection'strength (see Fig. \ref{fig:mmlt}). 

In the following subsections, we discuss the theory behind those two approaches and their impact on the stochastic excitation of acoustic modes.

\subsection{Critical magnetic field}
\label{sub:critical}
\par In non-magnetised fluids, \cite{schwarzschild_equilibrium_1906} derived an instability criterion for convection to begin: 
\begin{equation}
    \nabla - \nabla_{\rm ad} > 0,
    \label{eq:schwarz}
\end{equation}
where the temperature gradient is: 
\begin{equation} \nabla \equiv \frac{d \ln T}{d \ln P}, \end{equation}
and the adiabatic temperature gradient is: 
\begin{equation}\left.\nabla_{\mathrm{ad}} \equiv \frac{\partial \ln T}{\partial \ln P}\right|_s.\end{equation}
To take into account the effects of magnetic fields, \cite{gough_influence_1966} extended the previous criterion considering an ideal gas with a vertical magnetic field. They showed that convection exists if: 
\begin{equation}
    \nabla - \nabla_{ad} > \delta, 
    \label{eq:gough}
\end{equation}
where: 
\begin{equation}
    \delta = \frac{V_A^2}{V_A^2 + c_s^2}.
\end{equation}
$V_A$ is the Alfvén velocity introduced previously, and $c_s$ is the sound speed. Both Schwarzschild (Eq. \ref{eq:schwarz}) and Gough \& Tayler (Eq. \ref{eq:gough}) criteria make use of the energy principle of \cite{bernstein_ib_energy_1958}: they study the change of potential energy $\delta W$ of any fluid element under a small perturbation. Necessary and sufficient conditions for stability are obtained by minimizing $\delta W$ with respect to all the possible perturbations. 
One can then assess the critical magnetic field, above which convection is suppressed: 
\begin{equation}
    B_{crit} = \sqrt{\frac{\mu_0\rho c_{\mathrm{s}}^2 \left(\nabla -\nabla_{\mathrm{ad}}\right)}{1-\left(\nabla-\nabla_{\mathrm{ad}}\right)}}.
\end{equation}

\par \cite{gough_influence_1966}'s work has been pursued by \cite{newcomb_convective_1961}, who showed that a horizontal magnetic field does not influence convection. \cite{moreno-insertis_stability_1989}, \cite{macdonald_structural_2009}, \cite{macdonald_magnetic_2019} generalised Gough-Tayler criterion to take into account molecular mass variations $\mu$, non-ideal gas behavior, and radiation pressure. For simplicity, we will use Eq. (\ref{eq:gough}) derived by \cite{gough_influence_1966} in the following. In this framework, the critical inverse Alfvén number above which convection stops is: \\

\begin{equation}
    A_{crit} \equiv \frac{B_{crit}^2}{u_c^2 \mu_0 \rho_0} = \frac{-c_s^2 (\nabla - \nabla_{ad})}{u_c^2 (1 - \nabla + \nabla_{ad})}.
    \label{eq:acrit}
\end{equation}

Once again, as shown in Fig. \ref{fig:magnetised_convection}, the critical magnetic field approach fails to account for the progressive variations in the convective velocity $u_c$. To model more precisely the stochastic excitation of modes as seen in Section \ref{sec:scaling_laws}, one needs a more precise approach to magnetised convection. Here, we consider the Magnetic Mixing-Length Theory, as proposed by \cite{stevenson_turbulent_1979}.

\subsection{Magnetic Mixing-Length Theory (M-MLT)}

 Using a linear stability analysis of the convective instability, \cite{stevenson_turbulent_1979} derives scaling laws for the way the convective characteristic velocity and the convective characteristic wavenumber are modified by magnetic fields. This study was made under the hypothesis that the dominant convective mode is the one that carries the most energy \citep{malkus_heat_1997}.  \cite{stevenson_turbulent_1979} derives a modulation factor in each case: the convective velocity (resp. convective wavenumber) with magnetism $u_c$ (resp. $k_c$) is expressed with respect to the convective velocity (resp. convective wavenumber) without magnetism $u_0$ (resp. $k_0$): 
 \begin{equation}
\begin{aligned}
u_c = \tilde{U}(A) u_0 \\
k_c = \tilde{K}(A) k_0.  
\end{aligned}
\end{equation}

\begin{figure}
\centering
\includegraphics[width=0.5\textwidth]{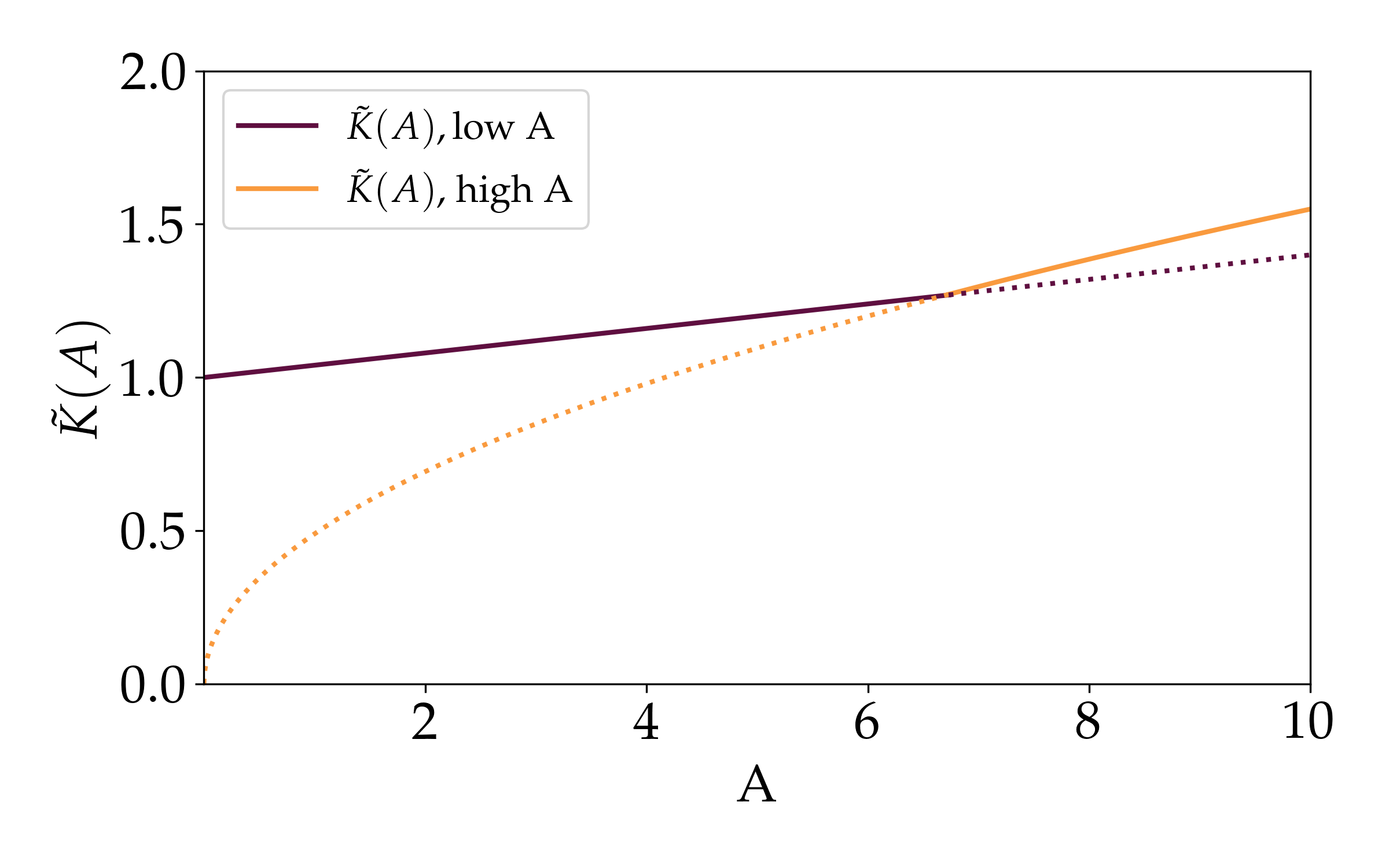}
\hfill
\includegraphics[width=0.5\textwidth]{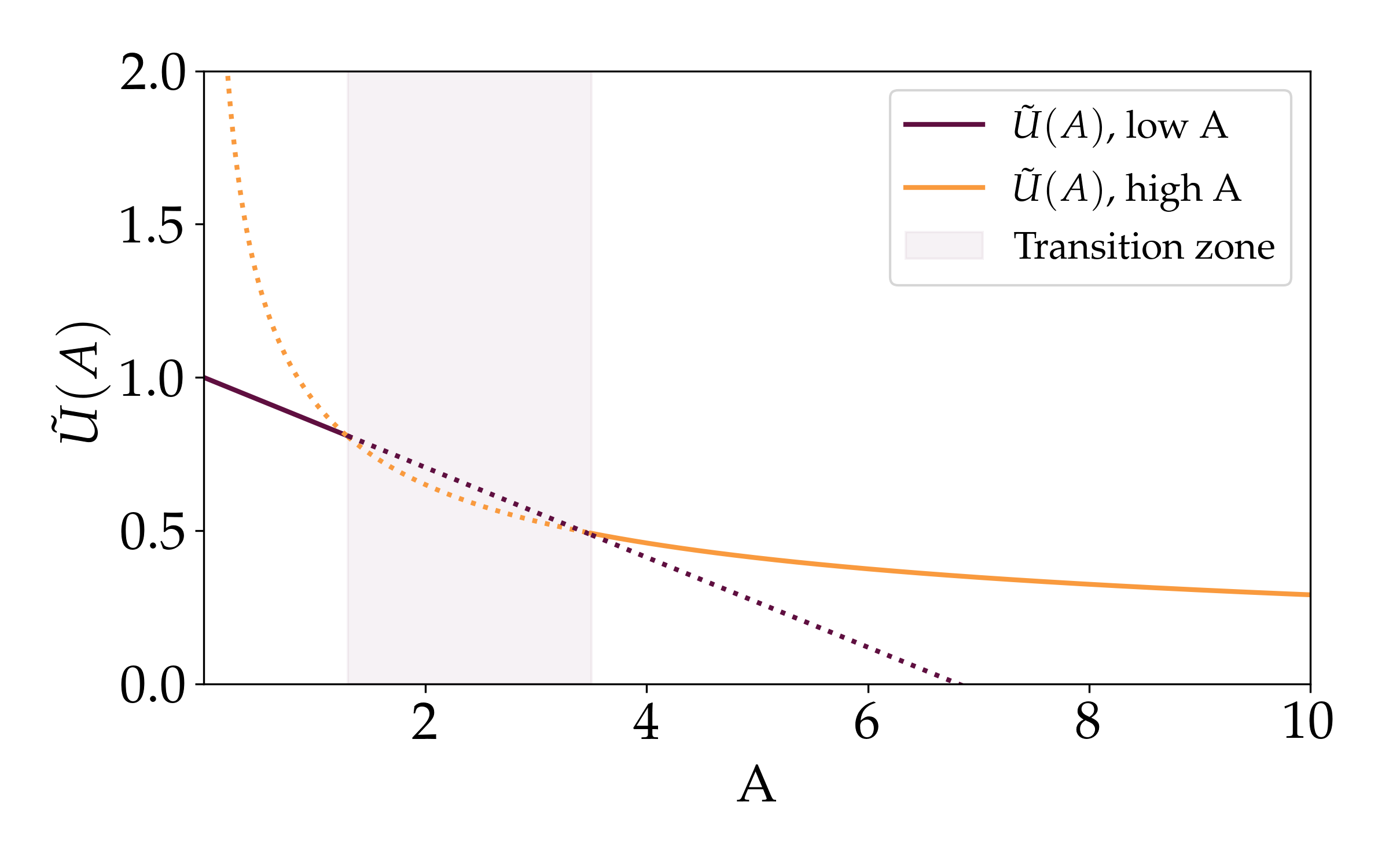}
\caption{Scaling laws given by Stevenson, for both the change in \textit{(Top)} wavenumber $\tilde{K}$ and \textit{(Bottom)} convective velocity $\tilde{U}$. The gray zone in the bottom figure, between $A = 1,28$ and $A=3,43$ shows the transition region where there is no clear distinction between the low-$A$ and the high-$A$ prescription. }
\label{fig:stevenson_mag}
\end{figure}

\cite{stevenson_turbulent_1979} derives scalings for the functions $\tilde{U}(A)$ and $\tilde{K}(A)$ for the asymptotic limits $A \ll 1 $ (low magnetic field) and $A \gg 1$ (high magnetic field). His prescriptions are detailed in Table \ref{tab:stevenson_mag}. These scaling laws are illustrated in Fig. \ref{fig:stevenson_mag}, for the low magnetic field and high magnetic field regimes. As expected, the convective velocity decreases when the magnetic field increases: the magnetic field has a stabilising effect on convection. Moreover, the convective wavenumber rises when the magnetic field is stronger. The size of the dominant eddy in MLT framework then diminishes, as previously illustrated in Fig. \ref{fig:stevenson_mag}.

\begin{table}[h!]
    \centering
    \renewcommand{\arraystretch}{2.5}
    \begin{tabular}{p{1.5cm}  p{2.5cm}  p{2.5cm}}
    \hline
     \hline
     & $A \ll 1$ &  $A \gg 1$ \\
     \hline
        $\tilde{U}(A)$ & $1-\frac{11A}{75}$   & $\frac{0,92}{\sqrt{A}}$\\
        
         $\tilde{K}(A)$ & $1 + \frac{A}{25}$   & $0,49 \sqrt{A}$\\
    \hline  
    \end{tabular}
    \caption{Values of the modulation factors $\tilde{U}(A)$ and $\tilde{K}(A)$ \citep{stevenson_turbulent_1979}.}
    \label{tab:stevenson_mag}
\end{table}

\par \cite{stevenson_turbulent_1979} also provides with prescriptions for rotating convection and set the ground to Rotating Mixing-Length Theory (hereafter R-MLT). This methodology has been extended adding the impact of diffusive processes in rotating convection \citep{augustson_model_2019}. In the low convective Rossby number regime, \cite{stevenson_turbulent_1979} scalings appear to hold well when compared to direct numerical simulations of rotating convection \citep{barker_theory_2014, vasil_rotation_2021, korre_dynamics_2021}, with interesting results for low-mass and massive stars internal structure and mixing for giant planets evolution \citep{michielsen_probing_2019, dumont_lithium_2021, fuentes_layer_2022}. Despite some numerical simulations of magnetised convection \citep[e.g.][]{hotta_breaking_2018}, no direct confrontation of \citeauthor{stevenson_turbulent_1979}'s prescriptions for M-MLT has been done yet. The physical approach for the rotating and the magnetic case being identical, one can expect robustness of the prediction for the Magnetic Mixing Length Theory.
\par
\begin{figure}
\centering
\includegraphics[width=0.5\textwidth]{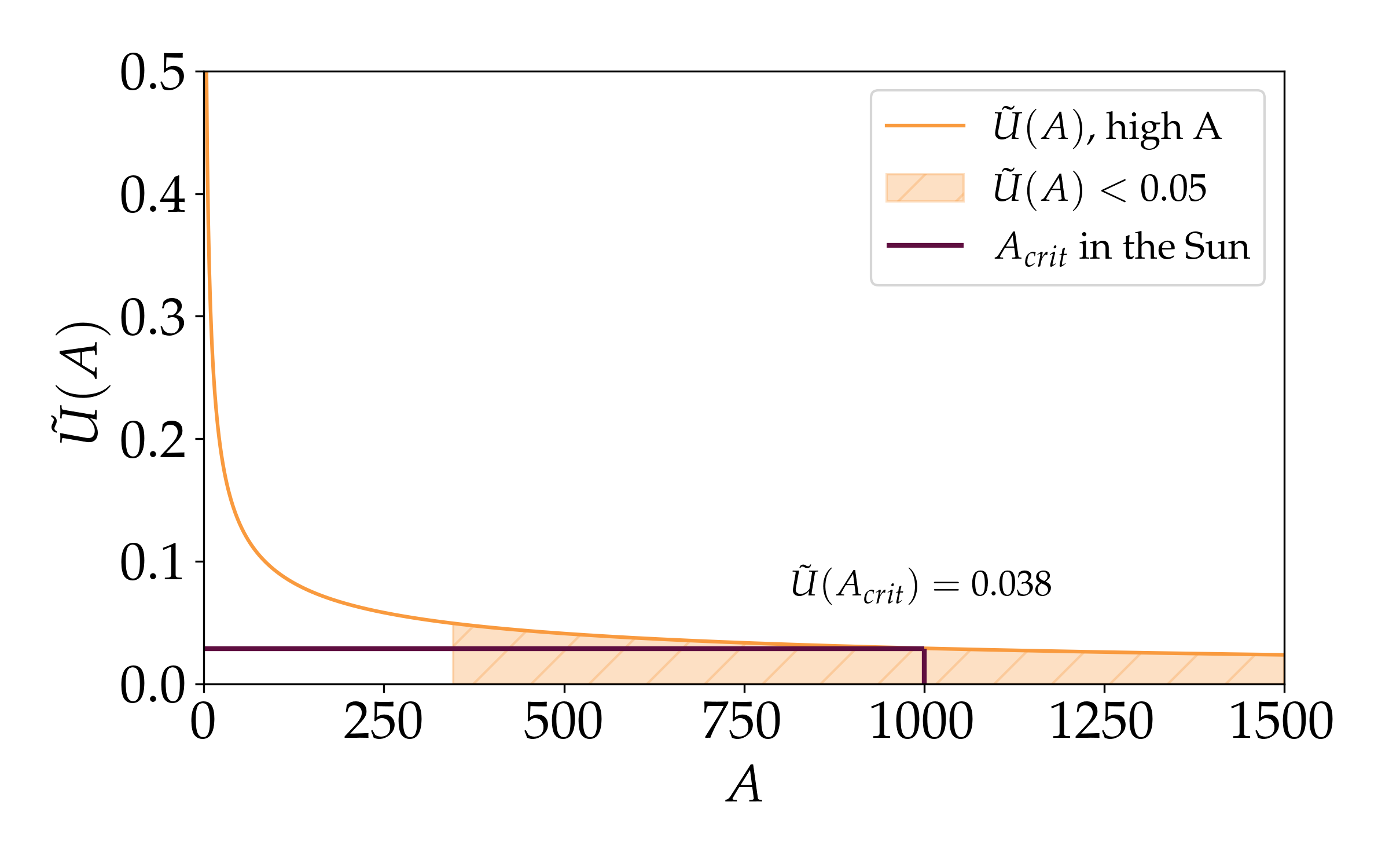}
\caption{Critical $A$ parameter in the Sun, confronted with Stevenson' scaling laws. The orange zone is where the convective velocity has diminished by more than $95\%$ with respect to the non-magnetised case. The purple line represents the position of the critical magnetic field in the Sun. }
\label{fig:critical_b_sun}
\end{figure}

 One way of checking Stevenson's prescription in M-MLT is to compare it with the critical magnetic field approach detailed in \ref{sub:critical}. To do so, we use the stellar structure and evolution code MESA \citep{paxton_modules_2011, paxton_modules_2013, paxton_modules_2015, paxton_modules_2018, paxton_modules_2019, jermyn_modules_2023} to generate a 1D stellar model of the Sun (see Appendix \ref{sec:appendix} for the inlist we used). At the top of the convective zone, we find a maximum $A_{crit} \sim 10^3$, with \cite{gough_influence_1966} stability criterion (Eq. \ref{eq:acrit}). According to Stevenson's prescriptions, the convective velocity has diminished by $97\%$ at this point: therefore, one can consider that convection is very weak compared to the non-magnetised case. The comparison is illustrated in Fig. \ref{fig:critical_b_sun}.
At first sight, \cite{stevenson_turbulent_1979}'scalings are thus consistent with the widespread critical magnetic field approach. It turns out to be an adequate prescription to use for the stochastic excitation of stellar oscillations in the presence of magnetism.

\section{Influence of magnetized convection on the resulting amplitude}

\label{sec:magnetized_convection_amplitude}

\subsection{Modification of the source and of the damping in magnetized convection}
\par Using \cite{stevenson_turbulent_1979} prescription of magnetised convection detailed in the previous section, we account for the modification of convection by magnetism for the driving and the damping of the oscillations. When there are no magnetic fields, the driving of the excitation boils down to the Reynolds-stresses source term, in non-magnetised convection. We note $\mathcal{S}_{0}$ the dominant excitation source without magnetism: 
\begin{equation}\mathcal{S}_{0} \sim \frac{-\rho_0 u_0^2}{\ell_{\rm osc}}.
\end{equation}
For reference, the symbols that will be frequently used from this section are listed in Table \ref{tab:notations}.
The magnetic field influences the stochastic excitation at different levels: 
\begin{enumerate}
    \item The magnetic Maxwell-stresses $\mathcal{S}_B$ also contribute to the excitation, along with the Reynolds-stresses $\mathcal{S}_R$.
    \item The convective velocity is modified by the magnetic field, both in the Maxwell-stresses $\mathcal{S}_B$ and in the Reynolds-stresses source term $\mathcal{S}_R$.
\end{enumerate}

When making use of M-MLT with the expressions of Table \ref{tab:stevenson_mag}, we find that the resulting source term with magnetic fields is modulated by a function $\tilde{\mathcal{S}}(A)$ when compared to the non-magnetised source term: 

\begin{equation}
    \mathcal{S}_{mag} \sim \mathcal{S}_{0} \underbrace{\tilde{U}(A)^2 \left(1 - \frac{A}{2 \tilde{U}(A)^2} \right)}_{\tilde{\mathcal{S}}(A)}.
\end{equation}
The modulation of the source term $\tilde{S}(A)$ is plotted in Fig. \ref{fig:scaling_source}, following the M-MLT scalings of \cite{stevenson_turbulent_1979} in Table \ref{tab:stevenson_mag}. In the low $A$ regime ($10^{-5}<A< 10^{-2}$, the magnetic field does not significantly lower the driving of the oscillations. From $A \sim 10^{-2} $ to $A \sim 1$, the source term is diminished when the magnetic field is higher: convection becomes lower due to the magnetic field. In the high $A$ regime ($A > 1$), the Maxwell-stresses become the dominant excitation source: $\tilde{\mathcal{S}}$ rises when $A$ increases. 

\begin{table}[h!]
    \centering
    \renewcommand{\arraystretch}{1.5}
    \begin{tabular}{p{2.5cm}  p{5cm} }
    \hline
     \hline
        $\mathcal{S}_R$ & Reynolds-Stresses source term   \\
        $\mathcal{S}_B$ & Maxwell-Stresses source term   \\
        $\mathcal{S}_{\rm mag} = \mathcal{S}_B + \mathcal{S}_R$ &  Total source term with $\boldsymbol{B} \neq 0$  \\
        $\mathcal{S}_0$ & Total source term with $\boldsymbol{B} = 0$   \\
        $\mathcal{D}_{mag}$ & Ohmic damping    \\
        $\mathcal{D}_{vis}$ & Turbulent viscous damping    \\
        $\mathcal{D}_{\rm mag} = \mathcal{D}_{\rm vis} + \mathcal{D}_{\rm ohm}$ &  Total damping term with $\boldsymbol{B} \neq 0$  \\
        $\mathcal{D}_0$ & Total damping term with $\boldsymbol{B} = 0$   \\
        $\langle \lvert A \rvert \rangle_0$ & Mean mode amplitude with $\boldsymbol{B} = 0$   \\
        $\langle \lvert A \rvert \rangle_{mag}$ & Mean mode amplitude with $\boldsymbol{B} \neq 0$   \\
        $\tilde{\mathcal{S}} = \frac{\mathcal{S}_{mag}}{\mathcal{S}_0}$ & Modulation of the source term with magnetic field   \\
        $\tilde{\mathcal{D}} = \frac{\mathcal{D}_{mag}}{\mathcal{D}_0}$ & Modulation of the damping with magnetic field   \\
        $\tilde{\mathcal{A}} = \frac{\langle \lvert A \rvert \rangle_{mag}}{\langle \lvert A \rvert \rangle_0}$ & Modulation of the amplitude with magnetic field  \\
    \hline  
    \end{tabular}
    \caption{Frequently used symbols in the stochastic excitation model.}
    \label{tab:notations}
\end{table}

\par Similarly, the damping terms are also modified because of the Ohmic diffusion and of the modification of the eddy viscosity since convective velocities and length scales intervening in its definition (Eq.\ref{eq:nu_turb}) are impacted by magnetic fields. We obtain:
\begin{equation}
    \mathcal{D}_{mag} \sim \mathcal{D}_0 \underbrace{\left(\frac{\tilde{U}(A)}{\tilde{K}(A)} + \frac{A^2 \mathcal{M}}{\mathcal{P}_m}\right)}_{\tilde{D}(A)}, 
\end{equation}
where $\mathcal{D}_0$ is the damping with no magnetic field, and $\tilde{\mathcal{D}}(A)$ is the modulation factor when adding the influence of a magnetic field. Note that the damping does not affect all the acoustic modes equally: it depends on the mode's velocity $u_{\rm osc}$. To give a trend of how damping is affected by magnetic fields, we consider a turbulent region where $\nu_t \sim \eta_B$, leading to $\mathcal{P}_m = 1$. In the Sun, the Mach number $\mathcal{M}$ reaches a maximum of $\mathcal{M} \sim 0.3$, at the top of the convective zone. It then leads to an upper limit for the damping. Under this simplification, the asymptotic expressions are detailed in Table \ref{tab:scaling_mag}, making again use of \cite{stevenson_turbulent_1979} asymptotic scalings in M-MLT. The damping of the oscillations slightly increases until $A \sim 1$: the Ohmic damping increases when the magnetic field is higher. However, the damping saturates in the high $A$ regime ($A > 1$). In this regime, the convection diminishes more and more with the increase of magnetic field, so the viscous damping decreases, which compensates for the increase of the Ohmic turbulent damping.

\begin{table*}[h]
    \centering
    \renewcommand{\arraystretch}{2.5}
    \begin{tabular}{p{1.5cm}  p{6cm}  p{6cm} }
    \hline
     \hline
     & $\tilde{\mathcal{S}}(A)$ & $\tilde{\mathcal{D}}(A)$ \\
     \hline
        $A \ll 1$ & $\left(1 - \frac{11 A}{75} \right)^2  \left(1 - \frac{A}{\left(1- \frac{11}{75 A^2}\right)^2}\right)$ & $\frac{11A}{75 (3A/25 + 3)} +1$  \\
    \hline
       $A \gg 1$ & $\frac{0,92^2}{A}\left(1 - \frac{A^2}{2 \cdot 0,92^2}\right)$ & $1.4508$ \\
        
    \hline  
    \end{tabular}
    \caption{Modulation of the source term of the oscillations ($\tilde{\mathcal{S}}$) and the damping ($\tilde{\mathcal{D}}$) in the asymptotic limits $A \ll 1$ and $A \gg 1$.}
    \label{tab:scaling_mag}
\end{table*}

\subsection{Mean squared amplitude with a magnetic field}
Knowing how the driving and the damping of the oscillations are affected by the magnetic field, one can give a tendency for the modification of the resulting modes amplitudes. Following Eq. (\ref{eq:amplitude_dirac}), we introduce $\langle \lvert A(t) \rvert^2 \rangle_{mag}$ the mean amplitude of the modes in the magnetised case, and $\langle \lvert A(t) \rvert^2 \rangle_{0}$, the mean amplitude in the non-magnetised case. As in the previous subsection, $\tilde{\mathcal{A}}$ accounts for the modulation of the mean modes amplitude by magnetic fields. Again with Eq. (\ref{eq:amplitude_dirac}), one has: 

\begin{equation}
    \tilde{\mathcal{A}} \equiv \frac{ \langle \lvert A(t) \rvert^2 \rangle_{mag} }{\langle \lvert A(t) \rvert^2 \rangle_0 } \propto \frac{\mathcal{S}_{mag}^2 \mathcal{D}_{0}}{\mathcal{S}_{0}^2 \mathcal{D}_{mag}} \sim \frac{\tilde{\mathcal{S}^2}(A)}{\tilde{\mathcal{D}}(A)}.
\end{equation}

\begin{figure}[h]
    \centering
    \includegraphics[width=0.5\textwidth]{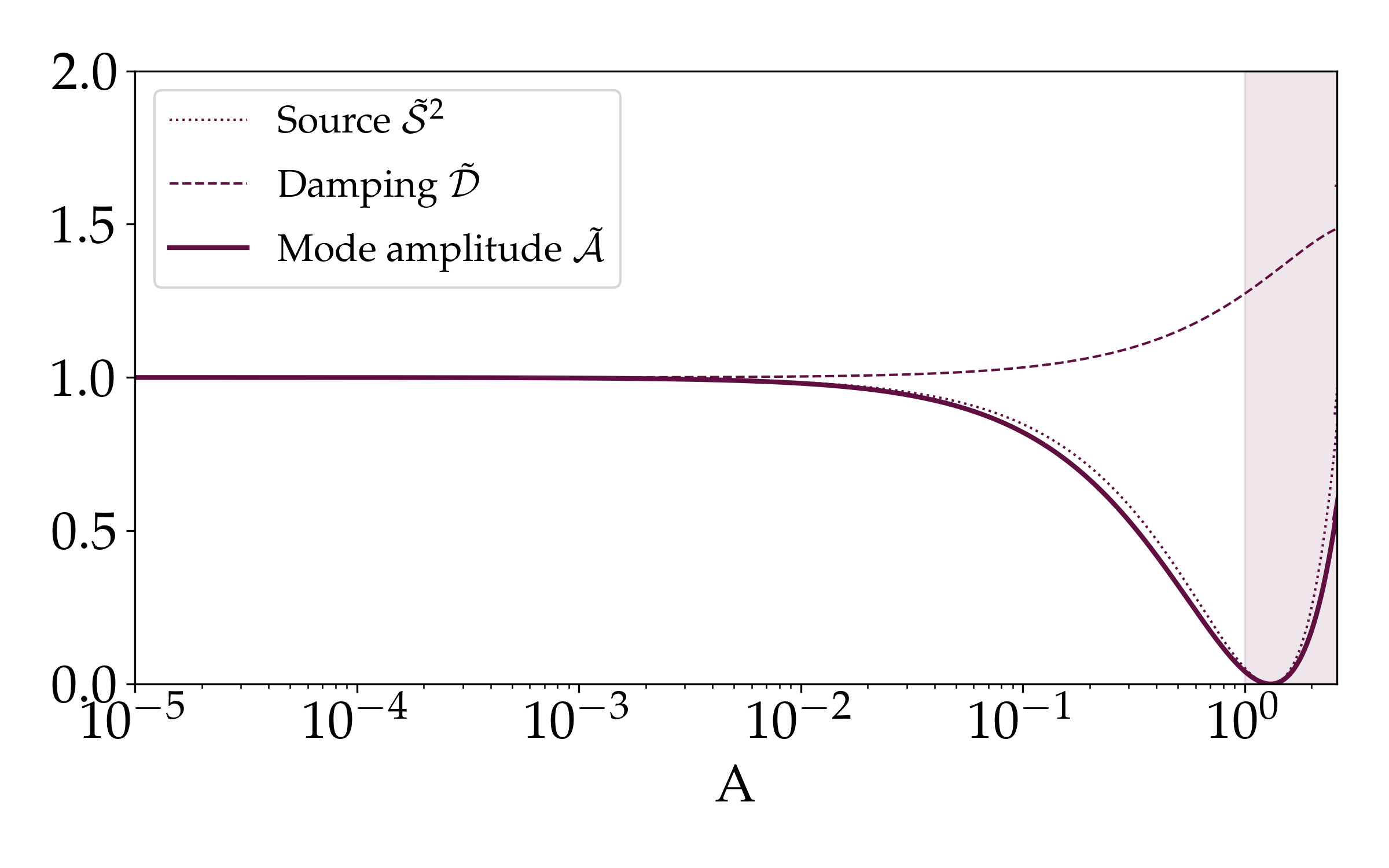}
    \caption{Scaling as a function of the inverse Alfvén number $A$. We clearly notice that the higher the magnetic field, the lower the source term, until $A= 1.28$. From this value, the magnetic Maxwell-stresses become the dominant excitation mechanism: the resulting injected power increases. However, the $A$ parameter in stars is below 1, so we expect to be always in the regime where the excitation is inhibited by the magnetic field.}
    \label{fig:scaling_source}
\end{figure}

Fig.\ref{fig:scaling_source} shows the resulting modulations of the modes amplitudes. Because of the distinct asymptotic regimes of magnetized convection in \cite{stevenson_turbulent_1979}, there is a discontinuity at $A = 1,28$. In this model, the damping slightly influences the modulation of the modes amplitude, but the overall tendency is dominated by the variations of the source terms $\tilde{\mathcal{S}}$. For low values of $A$, and more strongly between $A = 10^{-1}$ and $A = 1/2$ the magnetic field tends to inhibit the source of the stochastic excitation. Under $A = 10^{-4}$, there is nearly no change because the Maxwell-stresses source term is too low. For higher values of $A$, there is an increase in the excitation. As $ \mathcal{S}_B/\mathcal{S}_R \sim  A/2 $ so for $A > 2$, the Maxwell-stresses become higher than the Reynolds-stresses and the magnetic source term dominates the stochastic excitation. \\

\par This tendency is qualitatively in agreement with the observations of \cite{garcia_corot_2010}: the authors showed how the p-modes amplitudes were modulated along the magnetic cycle of a Sun-like star, HD49933. They used the frequency shifts induced by the magnetic field to assess the magnetic activity of the star. They observed that the stronger the magnetic field, the lower the modes amplitudes (and vice-versa). This is the tendency we predict as well, using these scaling laws. \\
The Sun's large-scale magnetic field at the surface is $B_{surf} \sim 4G-8G$, depending on the moment of the magnetic cycle). It corresponds to an inverse Alfvén number between $A = 7 \cdot 10^{-4}$ and $A = 1 \cdot 10^{-3}$: in this range, $\tilde{\mathcal{A}}^2(A)$ is close to one so the magnetic field slightly inhibits the excitation source, but one should not expect a significant change. \\

\section{Conclusion and perspectives}
\label{conclusion}
This work sheds light on the stochastic excitation of stellar oscillations when a magnetic field is present. It draws a first picture of the power injected into the modes considering both the magnetic source term that emerges, the Maxwell stresses source term, and the modification of convection by the magnetic field. After generalising the forced wave equation with a magnetic field, we provide scaling laws for the excitation source and the damping, making use of \cite{stevenson_turbulent_1979}'s prescriptions in the simplified framework of the Magnetic Mixing-Length Theory. Moreover, we used a 1-D Solar model to show that Stevenson's prescriptions for magnetized convection are in good agreement with the existing theory of a critical magnetic field above which convection is frozen.  We then demonstrated that the modes amplitudes tend to decrease with the magnetic field intensity for p-modes. It could explain the observations in \cite{garcia_asteroseismology_2019} where some modes amplitudes variations were detected following a magnetic cycle: the lower the magnetic activity, the higher the modes amplitudes and vice-versa. Therefore, it could explain why p-modes are not detected in stars with a high level of magnetic activity as this has been observed for solar-type stars observed by the Kepler mission \citep{mathur_revisiting_2019-2}. Such scaling laws should be confronted with the observations of the full \textit{Kepler} sample: an approach combining both 1-D stellar evolution modelling to access the inverse Alfvén number parameter, and observational data of p-modes will be relevant. Regarding the results presented in \cite{mathur_revisiting_2019-2}, they are presented as a function of the stellar rotation period and of the magnetic activity index, $S_{ph}$. In a near future, it would be important to establish the relationship between $S_{ph}$ and the magnetic field $B$, and in turn the inverse Alfvén number $A$ to compare the data with our theoretical prediction. To go further scaling laws, analytical work is ongoing to build on this theoretical formalism and render more precisely the contribution of different sources, using a spectral description of magneto-hydrodynamic turbulence. Such work will help model more precisely the stochastic excitation in the presence of magnetic field. 
\par Our work will be extended and adapted to tackle the stochastic excitation of gravity waves \citep{lecoanet_internal_2013}, as well as magneto-gravito-inertial waves \citep{mathis_low-frequency_2011, rui_gravity_2023, rui_asteroseismic_2024}.  These waves are excited at the base of the convective zone, as they are evanescent in the convective zone. Understanding their propagation and excitation is paramount, as they are one of the best candidates for the strong angular momentum transport needed in stellar radiative zones to reproduce the observed internal rotation revealed in stars by helio and asteroseismology \citep[see e.g.][]{schatzman_transport_1993, zahn_angular_1997, rogers_internal_2013}. It would in turn yield new insights into the rotational profiles and internal chemical mixing of rotating and magnetic stars. 
\par Finally, the magnetic field generated by the dynamo effect strongly depends on rotation \citep[e.g.][]{brun_magnetism_2017}: one must then study the combined impact of rotation and magnetic field on the stochastic excitation of stellar oscillations by turbulent convection. This would allow us to give more accurate predictions of the excitation of waves in both rotating and magnetised stars.

\begin{acknowledgements}
The authors are grateful to the referee for their detailed and constructive report, which has allowed us to improve this article.
L.B. and S.M. acknowledge support from the  European  Research Council  (ERC)  under the  Horizon  Europe program (Synergy  Grant agreement 101071505: 4D-STAR), from the CNES SOHO-GOLF and PLATO grants at CEA-DAp, and from PNPS (CNRS/INSU). While partially funded by the European Union, views and opinions expressed are however those of the author only and do not necessarily reflect those of the European Union or the European Research Council. Neither the European Union nor the granting authority can be held responsible for them.
\end{acknowledgements}

\bibliographystyle{aa}
\bibliography{bibliography.bib}

\appendix
\section{MESA Inlists for the Solar model}
\label{sec:appendix}
\begin{verbatim}
&kap
  ! kap options
  ! see kap/defaults/kap.defaults
  use_Type2_opacities = .true.
    Zbase = 0.016

/ ! end of kap namelist

&controls

      initial_mass = 1.0 

      ! MAIN PARAMS
      mixing_length_alpha = 1.9446893445
      initial_z = 0.02
      do_conv_premix = .true.
      use_Ledoux_criterion = .true.

      ! OUTPUT
      max_num_profile_models = 100000
      profile_interval = 300
      history_interval = 1
      photo_interval = 300

      ! WHEN TO STOP
      xa_central_lower_limit_species(1) = 'h1'
      xa_central_lower_limit(1) = 0.01
      max_age = 6.408d9

      ! RESOLUTION
      mesh_delta_coeff = 0.5
      time_delta_coeff = 1.0

      ! GOLD TOLERANCES
      use_gold_tolerances = .true.
      use_gold2_tolerances = .true.
      delta_lg_XH_cntr_limit = 0.01
      min_timestep_limit = 1d-1

      !limit on magnitude of relative change at any grid point
      delta_lgTeff_limit = 0.25 ! 0.005
      delta_lgTeff_hard_limit = 0.25 ! 0.005
      delta_lgL_limit = 0.25 ! 0.005

      ! asteroseismology
      write_pulse_data_with_profile = .true.
      pulse_data_format = 'FGONG'
      ! add_atmosphere_to_pulse_data = .true.

      ! rename the output directory
      log_directory = 'LOGS_SUN_Z_0.04'


/ ! end of controls namelist


&pgstar



/ ! end of pgstar namelist
\end{verbatim}

\end{document}